# Prediction of 2D ferromagnetism and monovalent europium ions in the EuBr/graphene heterojunctions

Haoyi Tan, [a] Guangcun Shan, [ab,*] Gianfranco Pacchioni[c]

[a] School of Instrumentation Science and Opto-electronics Engineering & Institute of Precision Instrument and Quantum Sensing, Beihang University, Beijing 100191, China.

[b] Department of Materials Science and Engineering, City University of Hong Kong, Kowloon Tong, Hong Kong SAR.

[c] Dipartimento di Scienza dei Materiali, Università degli Studi Milano-Bicocca, 20125 Milano, Italy

*Corresponding authors. **Email:** gcshan@buaa.edu.cn

**Abstract:**

Europium, one of the rare earth elements, exhibits +2 and +3 valence states and has been widely used for magnetic modification of materials. Based on density functional theory calculations, we predict the 2D EuBr/graphene heterojunctions to exhibit metallicity, huge intrinsic-ferromagnetism nearly 7.0 $\mu_B$ per Eu and the special monovalent Eu ions. Electron localization function (ELF), difference charge densities and Bader charge analyses demonstrate that there are cation-π interactions between the EuBr films and graphene, which explains the stability of these unusual heterojunctions. Graphene works as substrate to enhance the stability of EuBr monolayer crystals where EuBr plays an important role to imprint ferromagnetism and enhance metallicity in the heterojunctions. Monte Carlo simulations are used to estimate a Curie temperature of about 7 K, which, together with magnetic configurations, can be further modulated by external strains and charge-carrier doping. In general, our theoretical work predicts the properties of the novel 2D ferromagnetic EuBr/graphene heterojunctions, suggests the possibility of combining the 2D intrinsic-ferromagnetic metal halide crystals and graphene, and opens up a new perspective in next-generation electronic, spintronic devices and high-performance sensors.

## 1. Introduction

In recent years, the emerging of 2D magnetic materials has drawn much attention and set a new frontier of spintronics. In contrast to traditional magnetic materials, the reduced dimensionality makes 2D magnetic materials smaller, more susceptible to exhibit new physical phenomenon, which is intriguing and promising for both fundamental and practical interest. Nowadays, novel 2D magnetic materials include $Cr_2Ge_2Te_6$ [1], $CrI_3$ [2], $VSe_2$ [3], $Fe_3GeTe_2$ [4], hematene $Fe_2O_3$ [5], chromene $Cr_2O_3$ [6] and $CrBr_3$ [7]. However, the number of 2D magnets is still limited because of the suspension to 2D magnetism by strong spin fluctuations.

Rare earth elements belong to the group III B in the periodic table, including fifteen lanthanide elements (La-Lu) and two lighter elements, Sc and Y. 2D rare earth materials can exhibit intrinsic ferromagnetism because of their unpaired valence electrons, which makes them remarkable for magnetic modification of 2D non-magnetic Xenes (including graphene, silicene and germanene) [8]. The prospect to get ferromagnetism and extraordinary electronic properties combined and balanced in the same material is particularly appealing. In 2018, Tokmachev et al used the Eu element to functionalize silicene and therefore produced $EuSi_2$ silicene materials by molecular beam epitaxy. They found and AFM-to-FM transition going from bulk to ultrathin layers [9]. Later, Tokmachev et al and Sokolov et al followed the same approach to successfully obtain 2D ferromagnetic $EuGe_2$ and $EuC_6$ compounds, respectively, by using Eu to functionalize germanene or graphene [10,11]. In addition, graphene and EuS/EuO heterostructures were also explored and found to exhibit ferromagnetism [12,13]. Noteworthy, it is due to the formation of covalent bonds or van der Waals interactions that these structures can be obtained, but their synthesis is challenging.

Recently, Zhang et al successfully synthesized the novel 2D CaCl/graphene heterojunctions under ambient conditions by a simple saturated solution method [14]. Those CaCl crystalline thin films display metallicity, abnormal monovalent calcium ions and room-temperature ferromagnetism. Theoretical studies suggest that the formation of these unusual crystals is due to the strong cation–π interactions between the Ca cations and the aromatic rings in graphene [15], and the ferromagnetism observed experimentally originates from the edges and point defects of the 2D heterostructures, which means the intrinsic CaCl/graphene heterojunctions are nonmagnetic.

Inspired by the above studies, we predict a novel 2D ferromagnetic material, the EuBr/graphene heterojunction, based on density functional theory (DFT) calculations. Just like the CaCl crystals on graphene, 2D EuBr crystals are predicted to form a stable structure on graphene and have much stronger intrinsic ferromagnetism, nearly 7.0 $\mu_B$ per Eu. This amazing ferromagnetism is much stronger than what found in $CrI_3$ and $Fe_3GeTe_2$ due to unpaired 4f electrons of Eu. This is the first time that the cation-π interactions between the rare earth halide and graphene are predicted, including the possible existence of the special monovalent Eu ions. Our theoretical work may help to provide a new way to search the 2D ferromagnetic materials and imprint magnetism into 2D non-magnetic Xenes or their derivatives [16-18].

## 2. Computational details

We performed DFT using the Vienna ab initio simulation package (VASP). The exchange and correlation interactions were described within the generalized gradient approximation (GGA) using the Perdew-Burke-Ernzerhof (PBE) exchange-correlation functional [19-22]. The cut-off energy for the plane wave basis was set as 450 eV and a 21×21×1 Gamma centered k-mesh was used to sample the first Brillouin zone. In consideration of the strong correlation effectsof f-electrons, a Hubbard on-site Coulomb potential was set for Eu to U-J = 6.0 eV. The structures were optimized until the forces on each atom converged to 0.001 eV/Å. The accuracy of electronic self-consistency was set to be lower than $10^{-6}$ eV between the two electronic steps. The spin-orbit coupling effect was taken into account for all electronic and magnetic calculations. A vacuum layer of more than 20 Å is added to shield the interactions between periodic neighboring layers. Additionally, DFT-D3 corrections were included for van der Waals interactions [23,24].

## 3. Results and discussions

To search stable structures similar to the recently reported CaCl/graphene heterojunction, different metal halide crystals on graphene have been built and optimized based on DFT calculations, including three alkaline-earth halide crystals (Mg-X, Sr-X and Ba-X), seven transition metal halide crystals (Cr-X, Mn-X, Fe-X, Co-X, Ni-X, Cu-X and Zn-X) and two rare-earth halide crystals (Re-X and Eu-X), where X = F, Cl, Br and I. After optimization, we finally get –three nonmagnetic stable structures, MgBr/graphene, SrCl/graphene and BaCl/graphene, one antiferromagnetic, MnI/graphene, and one intrinsic-ferromagnetic EuBr/graphene heterojunction. The structure, magnetism and electronic properties of EuBr/graphene heterojunction are discussed below in details. The geometric structures, lattice constants and cohesive energy of the other heterojunctions are briefly discussed in figure S1 and table S1. The energy and magnetic moment of 2×2 supercells of MnI/graphene heterojunctions with different magnetic configurations are reported in table S2. It can be found that the lattice constants of the metal halide crystals and the valence states of the metals are factors that play an important role in the formation of stable 2D heterojunctions.

Figure 1(a) shows the 3×3 supercells of EuBr/graphene after geometrical optimization. The non-planar EuBr film exhibits a six-membered ring structure and perpendicularly stack above monolayer graphene. The whole heterojunction exhibits P3m1 symmetry with a lattice constant of 0.496 nm. The Eu-Br bond length is 0.310 nm, and the perpendicular distance between the Br and C atoms is 0.361 nm. The energy of 2×2 supercells of heterostructures with ferromagnetic (FM), antiferromagnetic (AFM) and ferrimagnetic (FIM) configurations has been calculated for comparison as shown in Figure 1(b) and Table 1(a).

Table 1(a) shows that the FM state is the ground state, with a magnetic moment of nearly 7.0 $\mu_B$ per primitive cell, i.e. per Eu ion. The magnetic moment is consistent with the $4f^7$ state of the Eu ions. The FIM ordering is the second most stable, 10.0 meV higher in energy than the FM state. The AFM1, AFM2 and AFM3 configurations have the same internal energy, which means all Eu ions are the same in lattice position and the heterojunctions have sixfold symmetry. We also calculated the magnetic configurations of EuBr/graphene heterojunctions using different U-J values for Eu to verify that the FM state is the energetically most stable, see table S3. If not differently mentioned, all the calculations reported below on EuBr/graphene are based on the FM ground state configuration. Table 1(b) lists the magnetic anisotropy energy of the primitive cells along different directions. The direction with the lowest energy corresponds to the easy magnetization axis of EuBr/graphene heterojunctions, and therefore magnetic moments are inclined to orient towards the +X direction.

Next, we try to evaluate the stability of EuBr/graphene heterojunctions. As shown in figure 2(a), there is nearly no imaginary frequency in phonon spectrum, which means the heterostructures are kinetically stable. Then we construct 5×5 supercells of EuBr/graphene heterojunctions and apply an ab initio molecular dynamics (AIMD) simulation to examine the thermal stability of heterojunctions through a slow heating progress with a heating rate of 0.15 K/fs. The above AIMD simulation is in a canonical ensemble (NVT) with a Nosé thermostat for temperature control. The critical temperature of structure failure is near 87 K, which is indicated by the sharp decline of energy in figure 2(b). We further calculated the elastic constants of EuBr/graphene heterojunctions, which have seven key parameters: $C_{11} = 117.521$, $C_{12} = 20.487$, $C_{13} = 0.029$, $C_{14} = 0.020$, $C_{33} = 0.003$, $C_{44} = 0.083$ and $C_{66} = 48.403$. According to the Born elastic stability criteria of rhombohedral (I) class crystals ($C_{11}>|C_{12}|$, $C_{44}>0$, $\frac{1}{2} C_{33}(C_{11}+C_{12})>C_{13}^2$ and $\frac{1}{2} C_{44}(C_{11}-C_{12})>C_{14}^2$) [25], EuBr/graphene heterojunctions are mechanically stable. The thermodynamic stability is tested by calculating the cohesive energy of EuBr/graphene heterojunctions, which is 6.98 eV/atom. For comparison, the cohesive energy of 2D graphene and isolated EuBr crystals is calculated to be 7.94 eV/atom and 2.69 eV/atom, respectively. The cohesive energy of 2D silicene and germanene is estimated to be 5.16 eV/atom and 4.15 eV/atom [26]. It is found that the presence of graphene as a substrate plays an important role for the stability of EuBr monolayer crystals. The formation of heterojunctions between the EuBr crystals and graphene slightly reduce the original thermostability of isolated graphene. But compared to 2D silicene and germanene, the EuBr/graphene heterojunctions are thermodynamically much more stable. The adhesion energy between EuBr crystals and graphene has also been calculated to estimate the stability of the heterostructures. The adhesion energy is about 1.05 eV/atom, which originates from the strong cation-π interactions and is one order of magnitude stronger than normal van der Waals interactions (about 0.1 eV/atom). In general, the above analyses demonstrate the stability of the EuBr/graphene heterojunctions.

In figure 2(c), the spin-dependent charge density is shown. The spin is mainly localized around the Eu ions, suggesting that the magnetism is mainly contributed by Eu 4f electrons. Magnetic coupling parameters of EuBr/graphene heterojunctions have been calculated based on Heisenberg model. $J_1$ and $J_2$ describe the magnetic exchange interactions between nearest-neighbor (NN) and second-NN Eu ions. The numbers of NN and second-NN Eu ions is 6 in both cases. The indirect interactions of third-NN Eu ions, which are described by $J_3$, have not been considered, because the distance of nearly 10 nm is too long and their influence is negligible. The calculated parameters are $J_1$ = -0.056 meV and $J_2$ = 0.010 meV, where the negative sign refers to a FM coupling and the positive sign to an AFM coupling. The Curie temperature is the transition temperature from ferromagnetic state to paramagnetic state, which means the magnetic directions of domains become random and disordered. Monte Carlo simulations have been applied to estimate the Curie temperature, which is about 7 K as shown in figure 2(d) [27,28].

In figure 3(a), ELF along the blue dotted line is analyzed in the top-right area of the panel, to determine the electronic localization and the type of bonding. The electrons are mainly localized around C and Br atoms, and an electron gas with an ELF value of about 0.5 exists around all atoms. The electron gas leads to good conductivity. Ionic bonds with an ELF value less than 0.5 are formed between the Eu and Br ions, and the interactions of C atoms in graphene, with an ELF value varying from 0.5 to 0.8, have the typical properties in between metallic and covalent bonds.

In figure 3(b), difference charge densities of the EuBr/graphene heterojunctions show how an electron transfer occurs from the Eu ions to Br ions and graphene. Bader charge analysis methods are further applied to quantitatively analyze the transfer of electrons in EuBr/graphene heterojunctions as shown in figure 4. It can be found that nearly 0.5 electrons transfer from Eu to Br ions to form ionic bonds, and another 0.5 electrons are transferred from the Eu ions to the aromatic carbon rings. These data indicate once more that it is the strong cation-π interactions and not just van der Waals interactions that are present between the EuBr film and monolayer graphene, which lead to the formation of these unusual EuBr/graphene heterojunctions with the special monovalent europium ions.

Considering the external stimulus can influence the magnetism of materials, we apply Monte Carlo simulations to further explore the influence of external strain and charge-carrier doping on the Curie temperature, as shown in figure 5(a and b). The charge-carrier doping is realized by directly increasing or reducing the number of electrons in EuBr/graphene heterojunctions. With the biaxial strains going from -5% to 5%, the Curie temperature of the heterojunctions decreases monotonically. When compressive strain reaches the highest value, the Curie temperature increases to 10 K; when tensile strain reaches the maximum, the Curie temperature decreases to 2 K. Upon charge doping, no matter if electrons or holes doping, the Curie temperature decreases, but hole doping affects the Curie temperature more than electron doping. Magnetic anisotropy energy of the EuBr/graphene heterojunctions as a function of external strain and charge-carrier doping is analyzed in figure 5(c and d). It is noteworthy that magnetic anisotropy energy can be reversed from in-plane to out-of-plane direction when a 5% compressive strain is applied on the EuBr/graphene heterojunctions.

The electronic properties of EuBr/graphene heterojunctions are studied by calculating the band structure. In figure 6(a and b), both spin-up and spin-down channels have the p-type Dirac point labeled by a red dotted circle. But when considering the spin-orbit coupling (SOC) effect, the Dirac point will open a small gap of about 0.11 eV as shown in figure 6(c). Projected band structure is studied to further reveal the contributions of different elements to electrical properties as shown in figure 6(d). It is the C $2p_z$ orbital that mainly contributes to Dirac point and electrons from Eu atoms that induce metallicity in this novel material. In contrast to graphene whose Dirac point is just at Fermi surface, the EuBr crystals in the EuBr/graphene heterojunctions play a role as donors to provide extra electrons and therefore shift the Fermi energy to higher values (The band structure of EuBr/graphene heterojunctions with different U-J set for Eu ions has also been calculated for comparison in figure S2.)

Finally, since CaCl/graphene heterojunctions have been successfully synthesized by Zhang et al through a simple saturated solution method under ambient conditions [14], we predict that the saturated solution method may also be appropriate for the synthesis of EuBr/graphene heterojunctions. Briefly, EuBr/graphene heterojunctions can be obtained by soaking the ultra-thin rGO membranes in a saturated $EuBr_2$ solution. In contrast to $EuGe_2$, $EuC_6$ or graphene/EuS (EuO) heterostructures, whose syntheses are based on molecular beam epitaxy, the synthesis of EuBr/graphene heterojunctions may be much easier and faster, which is also what we predict.

## 4. Conclusions

Based on DFT calculations, we predict the possibility to prepare novel 2D ferromagnetic EuBr/graphene heterojunctions with nearly 7.0 $\mu_B$ per Eu, in which Eu ions have an unprecedented monovalent oxidation state. The phonon spectrum, AIMD simulation, Born elastic stability criteria and cohesive energy analyses have been applied to confirm the kinetic, thermal, mechanical and thermodynamic stabilities of EuBr/graphene heterojunctions. By comparing the cohesive energy of isolated EuBr monolayer crystals and EuBr/graphene heterojunctions, it is found that graphene plays an important role as substrate to enhance the stability of EuBr crystals. By applying Monte Carlo simulations, the Curie temperature of EuBr/graphene heterojunctions is estimated to be about 7 K, which together with magnetic configurations can be further modulated by external strain and charge-carrier doping. Analysis of the electronic structure based on ELF, difference charge densities, and Bader charge analyses show the strong cation-π interactions between the Eu ions and aromatic rings of graphene that induce the formation of such novel 2D heterojunctions and give rise to the special stability of monovalent Eu ions. By analyzing the band structure, EuBr crystals not only imprint ferromagnetism into 2D non-magnetic graphene, but also act as donor to provide extra electrons and enhance the metallicity in heterojunctions. In conclusion, this work demonstrates theoretically the possibility of combining a 2D intrinsic-ferromagnetic metal halide crystal and graphene, and paves the way towards other potential materials for spintronics and the induction of ferromagnetism in 2D non-magnetic Xenes or their derivatives.

**Conflicts of interest:**

There are no conflicts to declare.

**Acknowledgements**

This work was financially supported by National Natural Science Foundation of China (No.21771017) and also the Fundamental Research Funds for the Central Universities.

**Figures**

Figure 1. The ball-and-stick models of EuBr/graphene heterojunctions. (a) Geometric structures of 3×3 supercells and (b) heterojunctions with different magnetic configurations. (Red arrows represent for spin up and blue arrows represent for spin down)

Figure 2. (a) Phonon spectrum, (b) internal energy as the function of temperature, (c) spin-dependent charge density (represented by the yellow ball regions) and (d) average magnetic moment as the function of temperature.

Figure 3. (a) ELF and (b) difference charge densities of the EuBr/graphene heterojunctions. Yellow regions indicate gaining electrons and cyan regions indicate losing electrons.

Figure 4. Bader charge analysis. (Pink marked atomic information is from benzene ring just under Eu ions)

Figure 5. Curie temperature and magnetic anisotropy energy of the EuBr/graphene heterojunctions as a function of (a and c) external strain and (b and d) charge-carrier doping.

Figure 6. Band structure of the EuBr/graphene heterojunctions for the (a) spin up and (b) spin down channels without SOC effect, (c) band structure and (d) projected band structure with SOC effect. The insets in (b and d) represent the high symmetry q point path in the Brillouin Zone and the contribution of C $2p_z$ orbit to Dirac point, respectively.

**Figure 1**

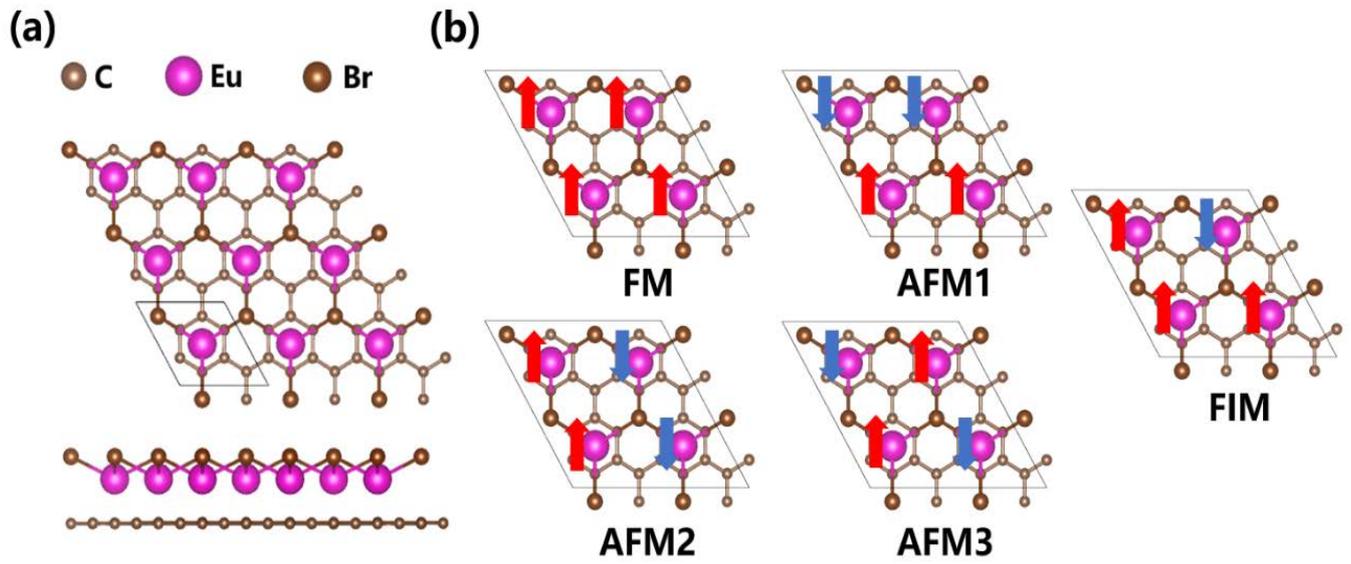

**Figure 2.**

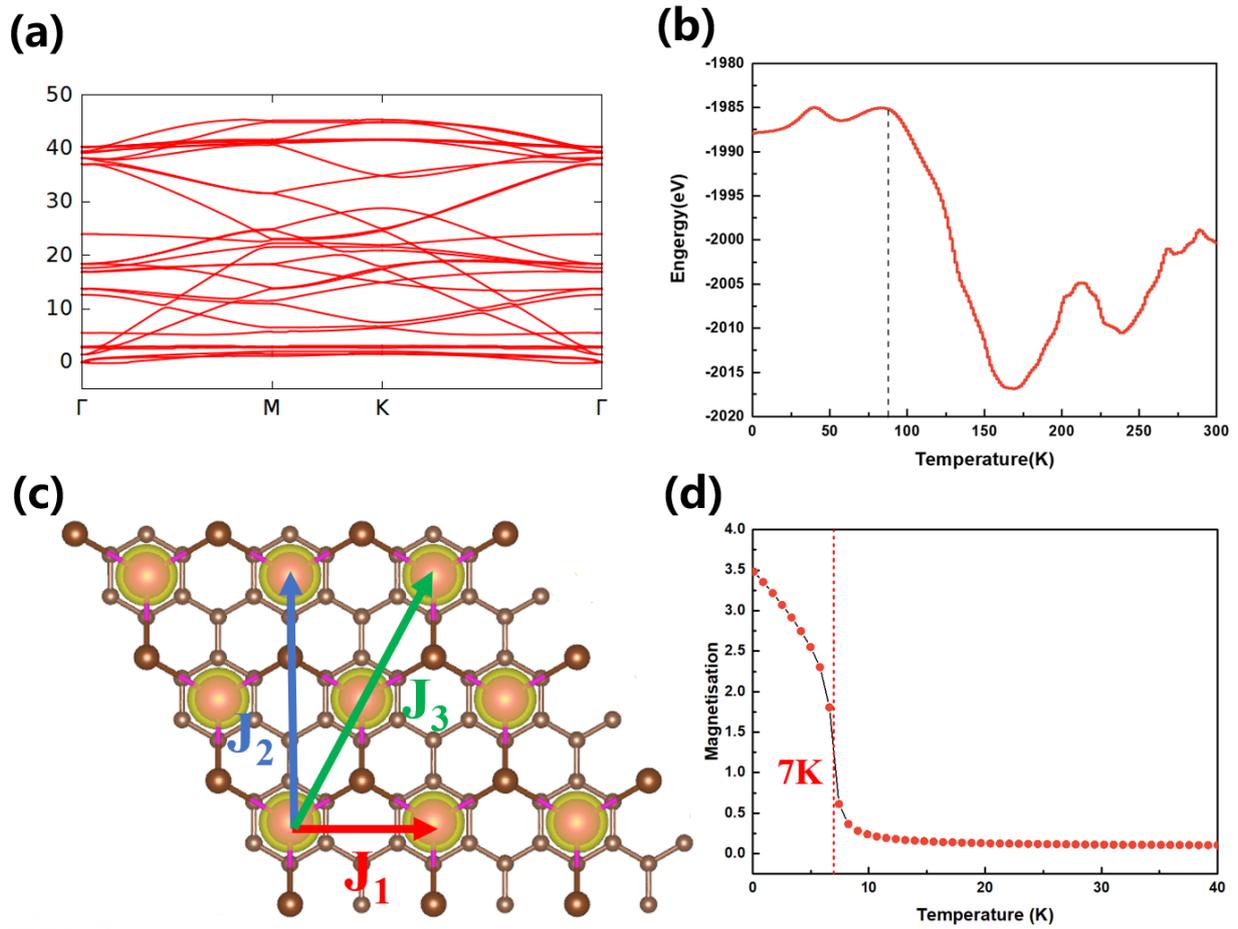

**Figure 3**

(a)
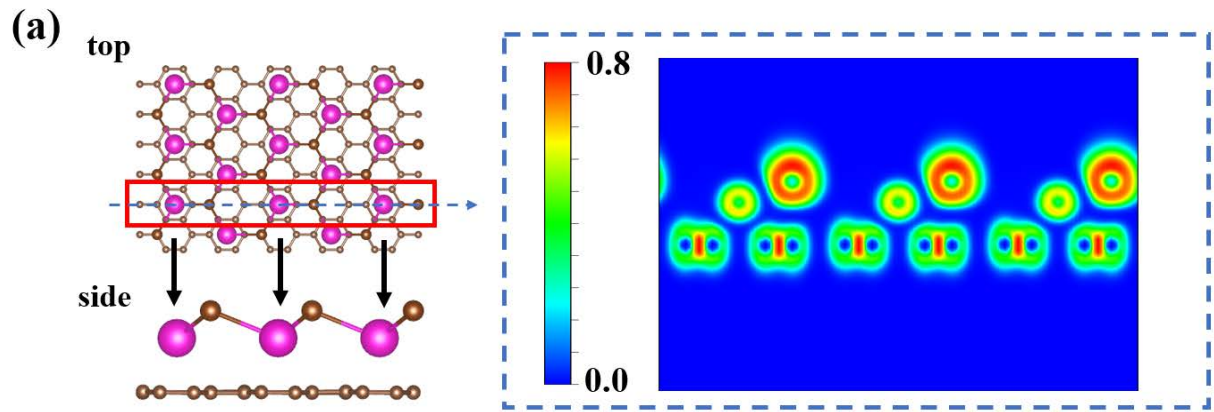

(b)
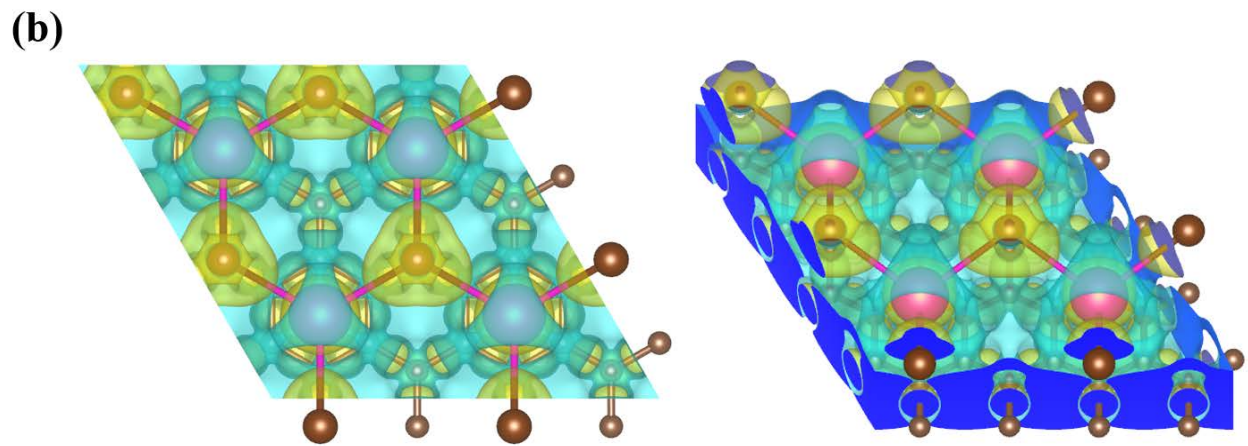

**Figure 4**

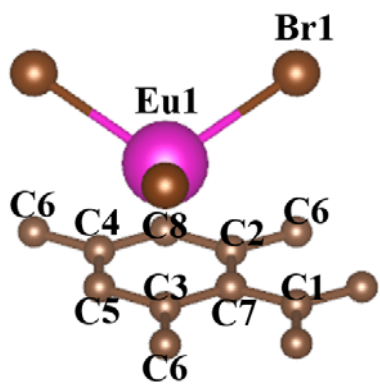

| Atom Labeling | Initial Charge | Final Charge | Transferred Charge |
|---|---|---|---|
| C1 | 4.00 | 3.99 | -0.01 |
| C2 | 4.00 | 4.22 | 0.22 |
| C3 | 4.00 | 4.01 | 0.01 |
| C4 | 4.00 | 4.11 | 0.11 |
| C5 | 4.00 | 4.15 | 0.15 |
| C6 | 4.00 | 4.01 | 0.01 |
| C7 | 4.00 | 4.02 | 0.02 |
| C8 | 4.00 | 4.04 | 0.04 |
| Eu1 | 17.00 | 15.97 | -1.03 |
| Br1 | 7.00 | 7.48 | 0.48 |

**Figure 5**

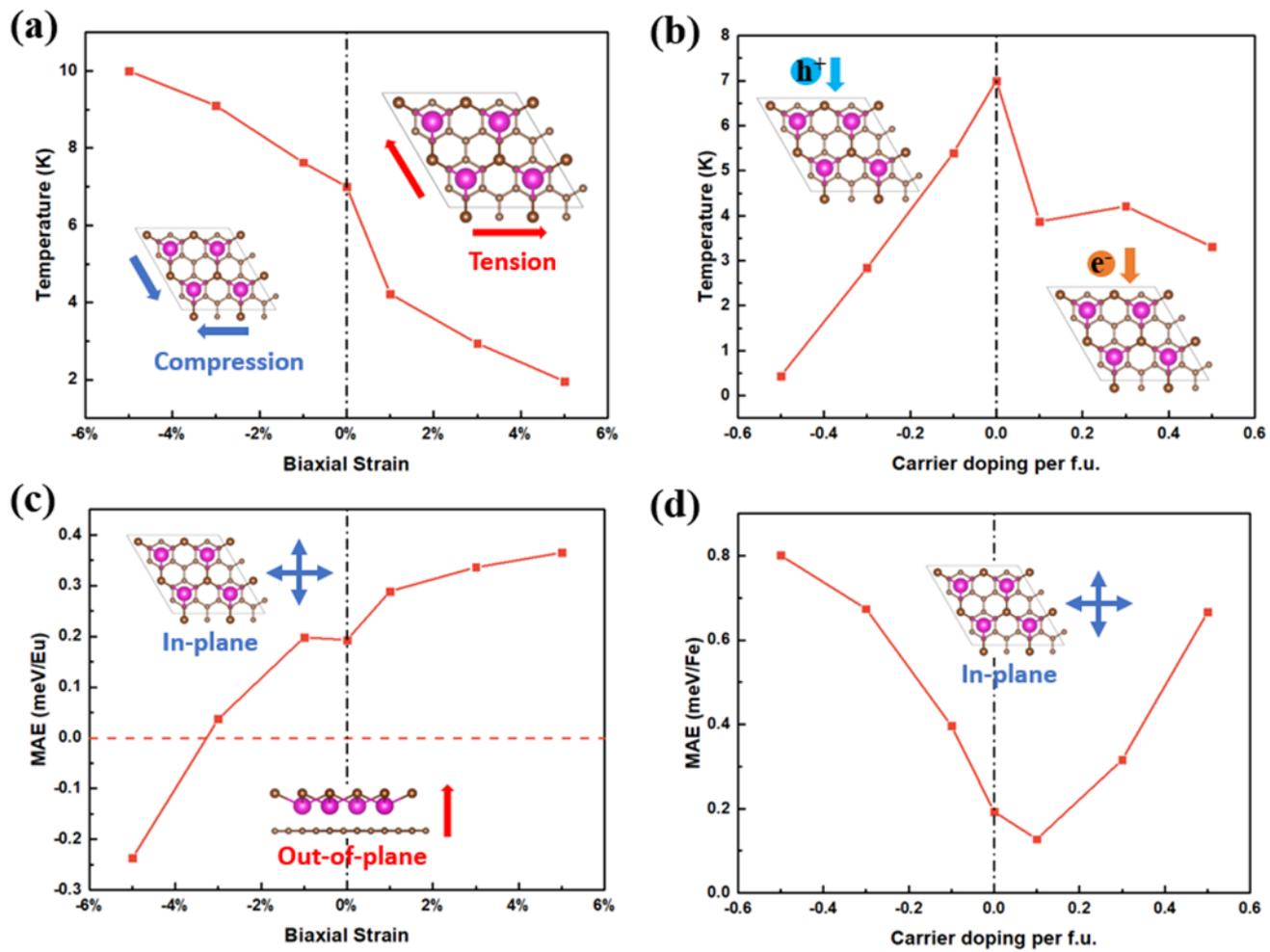

**Figure 6**

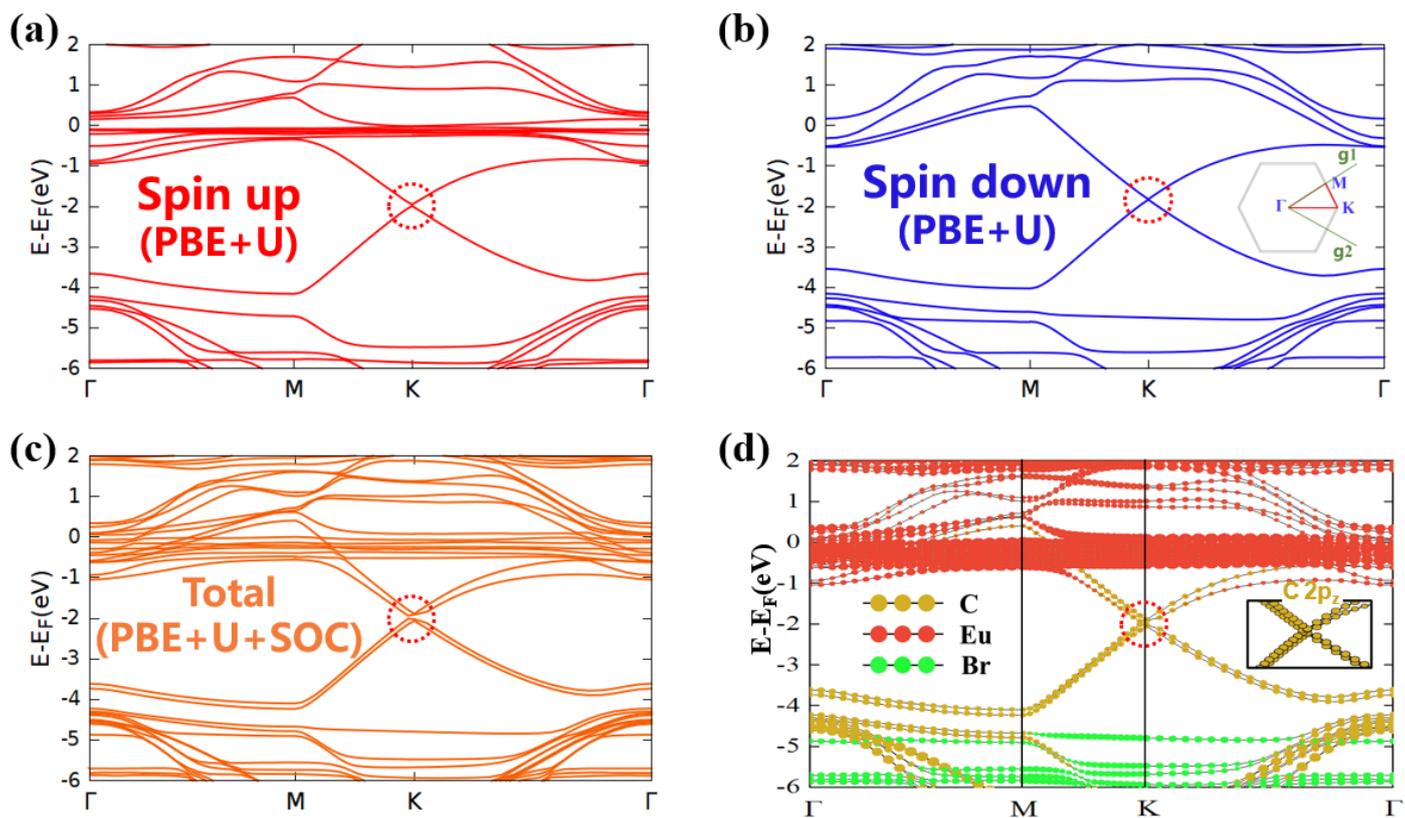

**Table 1. (a) Energy and magnetic moment of 2×2 supercells of EuBr/graphene heterojunctions with different magnetic configurations and (b) magnetic anisotropy energy of primitive cells.**

(a)

| Magnetic Configurations | Energy (eV) | Magnetic Moment ($\mu_B$) |
|---|---|---|
| FM | -358.19317 | 28.0 |
| AFM1 | -358.18051 | 0.0 |
| AFM2 | -358.18051 | 0.0 |
| AFM3 | -358.18051 | 0.0 |
| FIM | -358.18316 | 14.0 |

(b)

| Magnetic Anisotropy | Energy (eV) | Energy Difference (meV) |
|---|---|---|
| +X | -89.54976 | 0 |
| -X | -89.54779 | 1.98 |
| +Y | -89.54921 | 0.55 |
| -Y | -89.54921 | 0.55 |
| +Z | -89.54957 | 0.19 |
| -Z | -89.54795 | 1.81 |